\documentclass[aps,showpacs,preprintnumbers,amsmath,amssymb,superscriptaddress,floatfix,a4paper,nofootinbib,11pt,floatfix,nofootinbib,onecolumn]{revtex4}
\usepackage[pdftex]{graphicx}
\usepackage{bm}
\usepackage{float}
\usepackage{color}
\usepackage{dcolumn}    
\usepackage[spanish,english]{babel}
\usepackage{bm}     
\usepackage{bbm}       
\usepackage{amssymb}  
\usepackage{amsmath}
\usepackage{latexsym}
\usepackage{ifthen}
\usepackage{caption,subfig}
\usepackage{enumerate}
\usepackage{url}
\usepackage{caption,subfig}
\usepackage{amsopn}
\usepackage{hyperref}
\usepackage{amsfonts}
\usepackage{multirow}
\usepackage{array}
\usepackage{booktabs}
\usepackage{rotating}
\DeclareMathOperator{\eln}{\ell n}

\usepackage{amsmath}

\usepackage{ulem}
\begin{document}

\title{Unraveling the vertical motion of {\it Dipterocarpus alatus} seed using Tracker}

\author{Thammarong Eadkong}
 \affiliation{School of Science, Walailak University, Nakhon Si Thammarat 80160, Thailand}

\author{Pimchanok Pimton}
\affiliation{School of Science, Walailak University, Nakhon Si Thammarat 80160, Thailand}
\affiliation{Research Group in Applied, Computational and Theoretical Science (ACTS), Walailak University, Thasala, Nakhon Si Thammarat, 80160, Thailand}

\author{Punsiri Dam-O\footnote{Corresponding author: dpunsiri@mail.wu.ac.th}}
\affiliation{School of Science, Walailak University, Nakhon Si Thammarat 80160, Thailand}
\affiliation{Research Group in Applied, Computational and Theoretical Science (ACTS), Walailak University, Thasala, Nakhon Si Thammarat, 80160, Thailand}
        
\author{Phongpichit Channuie}
\affiliation{School of Science, Walailak University, Nakhon Si Thammarat 80160, Thailand}
\affiliation{Research Group in Applied, Computational and Theoretical Science (ACTS), Walailak University, Thasala, Nakhon Si Thammarat, 80160, Thailand}
\affiliation{College of Graduate Studies, Walailak University, Thasala, Nakhon Si Thammarat, \\80160, Thailand}
  
\date{\today}

\begin{abstract}
\subsection*{Abstract}
We perform analytical and experimental investigation of the vertical motion of {\it Dipterocarpus alatus} seed, locally called Yang-na in Thailand. In this work, we assume the drag forces exerting on the Yang-na seeds depend only on the velocity. We derive equations of motion (EoMs) to physically parametrize the vertical motion of the seed and analytically solve to obtain the exact solutions. Interestingly, we observe that our predicted solutions are in agreement with the experimental data. More precisely, the entire trajectory of the falling seed of Yang-na can be described by our predicted solutions. We also determine terminal velocity of the seeds. Remarkably, this work reasonably proves that seed dispersal characteristics of Yang-na is inherently straight downward. Finally, we believe that our achievement will be valuable to the large community of
STEM/STEAM education to promote an understanding in the topic integrating mathematics, physics,
biology, art and technology. Our framework constitutes learning model to improve the ability of creative thinking, analytical thinking and problem solving skills on the concept of forces and motion
applicable from high school to college levels. \\

Keywords: AutoTracker; Dipterocarpus alatus seed; Drag and Lift forces


\end{abstract}

\maketitle

\section{Introduction} 
\label{intro}
{\it Dipterocarpus alatus}, or Yang-na in Thai language, is one of the tree species in Dipterocarpaceae family. Yang-na trees can be used in house construction, plywood production and chock wood for train. Its rasin, a minor forest product, is a coating material to conserve wooden houses and boats. Because of the multi-purposes of Yang-na trees, they were intensively used. Some species in Dipterocarpaceae family are nearly extinct \cite{VL} in the countries of tropical lowland and rainforest (e.g. North America, Africa and Southeast Asia). In Thailand, since 1980s, the population of Yang-na has been decreasing \cite{TS}. In order to save this species for the next generation, Thai public has expressed the protection of Yang-na through law approach and scientific research. In a viewpoint of physicists, a seed of Yang-na can present its interesting motion. Yang-na trees have evolved seeds with two unique curvature wings, enabling them to be carried away by the wind when falling \cite{KT}. If the dynamics of the falling seed is understood, the location of the germinating seed will be predicted.  

The process of the seed falling starts at 20 upto 40 meter-height of the tree. After a few seconds, the whole seed flips upside down to let the wings position at the upper part of its seed, and from this moment, the wings fully undergo air resistance, where the seed begins to rotate towards the ground. Rotation is a natural mechanism enabling the falling seed to slow down its descent, and a sense of rotation is generally dictated by right-left handedness \cite{KT}. Due to the intrinsic natural design, Yang-na seeds possess a geometrical right-left asymmetric wing morphology, allowing the seeds to rotate and disperse over the area in their surrounding. However, the seeds of Yang-na are larger and heavier compared with other species in the same tree family \cite{SS}. As a result, the dispersal of Yang-na seeds is reported to be poor, mostly around the mother tree \cite{JR,SA}. 

The complex motion analogous to the seeds of Yang-na is also observed in those of maple, mahogany and {\it Triplaris caracasana}. A number of research groups conducted the studies to explain the motion of falling wing seeds. In 2009, Celso and Pedro \cite{CL} discussed about the vertical and rotational motion of the tri-wing seed. The vertical motion starting with the non-linear transient and followed by the final uniform terminal speed. They theoretically predicted the angular speed of the rotating seed and found its good agreement with their experiment. 

Here, we use Tracker video analysis software to explore the vertical motion of the Yang-na seed and to test an underlying theory. Tracker is a free and open source software and can be downloaded at www.opensourcephysics.org \cite{DB}. It is widely used for analysing the movement of objects in a recorded video, especially for Physics education. For example, Eadkhong {\it et.\,al.} (2012) used Tracker to analyse the rotational motion of a spinning cylindrical plate to determine the moment of inertia \cite{TE}. Wee {\it et.\,al.} (2012) reported the use of Tracker to analyse the projectile motion of a tossing ball \cite{LKW}. Kinchin (2016) applied Tracker to analyse the movement of a string pendulum to demonstrate the simple harmonic motion \cite{JK}. More recently, Suwarno (2017) used Tracker to analyse the rotating movement of the fidget spinners \cite{DUS}. In these experiments, the position as a function of time of the moving object is video captured and, by using functions from the Tracker's library, the phenomena can be simulated. Tracker allows to compare the graphs of the video-based experimental data with the simulations. 

In this work, Yang-na seed samples were collected within the campus area of Walailak University, Nakhon Si Thammarat, Thailand. We experimentally drop the seed sample from a fixed height and demonstrate that a vertical motion of a seed can be completely described by assuming that the drag forces exerting on the seed depend only on the velocity. In Sec.(\ref{sec2}), we start by considering a free-body diagram of the seed and derive equations of motion (EoMs) to physically parametrize the vertical motion of the seed, mathematically. We then analytically figure out exact solutions of the EoMs describing the its behavior. In Sec.(\ref{sec3}), we display our experimental setup in order to obtain real data. In Sec.(\ref{sec4}), we perform experimental analysis. In this section, we also confront our predicted solutions with the  experimental data obtained by the use of Tracker. Finally, we conclude our finding in the last section. 

\section{Background Equation and the analytical solutions}
\label{sec2}
\begin{figure}[H]
	\begin{center}
		\includegraphics[width=0.33\linewidth]{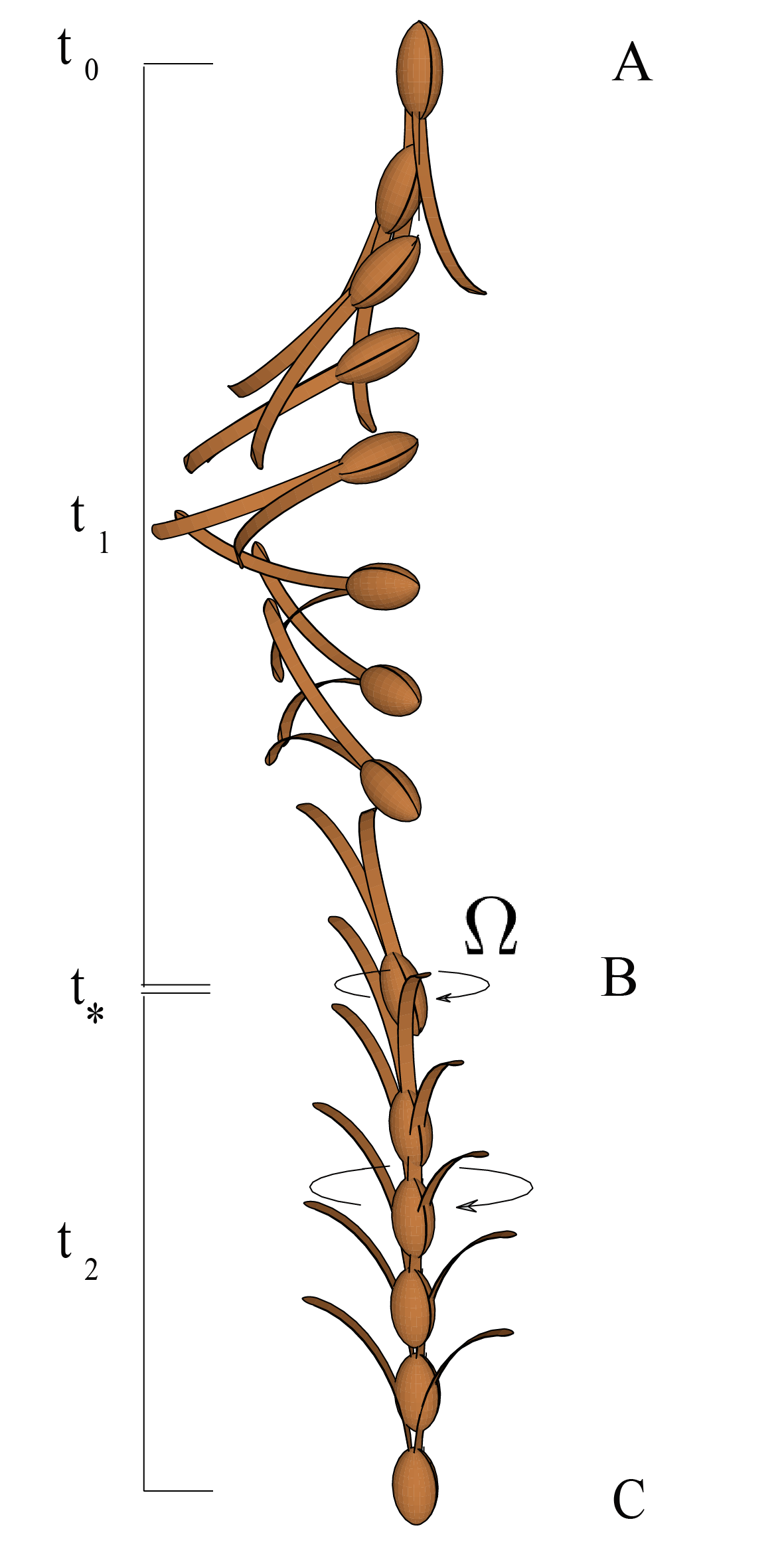}
		\caption{Schematic diagram illustrating the morphology of Yang-na seed during vertical motions. A flight sequence of a seed showing its center of gravity situated along the vertical axis from A to C at time $t_{0}$ to $t$. At B, the seed starts to spinning. An angular frequency is given by $\Omega$.} \label{sep}
	\end{center}
\end{figure}

In this work, we consider the objects ({\it Dipterocarpus alatus} seeds) which are falling under the influence of not only gravity but also the drag forces. We assume the later are functions of the velocity. Schematic diagram illustrating the morphology of Yang-na seed during vertical motions is displayed in Fig.(\ref{sep}). From the free-body diagram, we can deduce the forces exerting on the object by two intervals, i.e. $t_{0}\leq t_{1} \leq t_{*}$ and $t_{*}\leq t_{2} \leq t$. The first interval represents the time of flight just after the falling ($t_{0}$) to just before the auto-rotation ($t_{*}$); whilst the second one is during its autorotation. We call the first interval \lq\lq falling and flipping\rq\rq and the second one \lq\lq auto-rotation\rq\rq.

\subsection{Falling and flipping}
 
Let us first consider the first interval, i.e., falling and flipping, and assume the forces ${\vec F}_{1}={\vec F}_{D}+{\vec F}_{g}$. Therefore its equation of motions takes the form:
\begin{eqnarray}
{\vec F}_{D}+{\vec F}_{g} = m{\vec a},\label{str0}
\end{eqnarray}
where $m$ represents the mass of the object, ${\vec a}$ is the gravitational acceleration, ${\vec F}_{D}$ and ${\vec F}_{g}$ are drag and gravitational forces, respectively. Notice that when neglecting the drag forces, the object is free falling. In this work, we assume that the center of gravity of the flight is always aligned along the vertical line illustrated in Fig.(\ref{sep}). In order to describe the motion, we consider that the drag force takes the form 
\begin{eqnarray}
F_{D,y} = k'_{1}(v^{2}+u^{2}).\label{str00}
\end{eqnarray}
where a parameter $k'_{1}$ is the drag coefficient, $v$ and $u$ are the vertical and horizontal velocity components of the falling seed, which are time-dependent quantities. In a close experimental room without air flow, the parameter $u$ is equivalent to zero and the relative velocity of the seed relative to the air is then corresponded to the speed of the falling seed. Eq.(\ref{str00}) reduces to $F_{D,y} = k'_{1}v^{2}$. The Reynolds number in this experiment is estimated to be of the order of $10^{3}$ in which drag coefficient is relatively constant \cite{JDbook}. Therefore, we can write
\begin{eqnarray}
ma_{y_{1}} = -mg+k'_{1}v(t)^{2}, \label{str1}
\end{eqnarray}
where for convenience we have considered the modulus of the vector quantities that are functions of time, $t$, explicitly. In this work, we assume that an acceleration in $x$-direction can be approximately omitted. In terms of a displacement in $y$-direction, we can simply express the above equation to yield
\begin{eqnarray}
m\frac{d^{2}y_{1}(t)}{dt^{2}} = -mg+k'_{1}\left(\frac{dy_{1}(t)}{dt}\right)^{2}.\label{str12}
\end{eqnarray}
At the initial point, $t=t_{0}$, only gravitational force presents. It is possible to quantify the solution of Eq.(\ref{str12}) and it can be analytically solved to obtain
\begin{eqnarray}
y_{1}(t)= y_{*,1}-\frac{m}{k'_{1}} \eln \left(\cosh \left(\frac{\sqrt{g\, k'_{1}}\,t}{\sqrt{m}}\right)\right),\label{sol1}
\end{eqnarray}
where $t_{0}\leq t\leq t_{*}$. Note that the drag coefficient $k'_{1}$ will be determined by fitting the above predicted solution given in Eq.(\ref{sol1}) with the experimental data. Note that $t_{0}$ is a time at which the seed starts falling, while $t_{*}$ denotes time that the seeds just start to take an auto-rotation.   

\subsection{Auto-rotation} 

As mentioned in Ref.\cite{KT}, an efficient physical mechanism that enables slowing down the descent is rotation reducing the translational kinetic energy. In this second stage of the motion, the lift and torque are generated by rotating wings. Therefore, the blade element model can be used to describe the autorotating seeds \cite{Norb,Aki,Lee}. Since the geometric curvature of the wing of Yang-na seed displays a skewed blade element, the leading edge vortex is stably developed during the descend of the seed. Additionally, it has been elucidated that the leading edge vortex is responsible for raising both lift and drag forces during the autorotating falling \cite{DL2009,JS2019,JJ2019,TM1979}.
\begin{figure}[H]
	\begin{center}
		\includegraphics[width=0.4\linewidth]{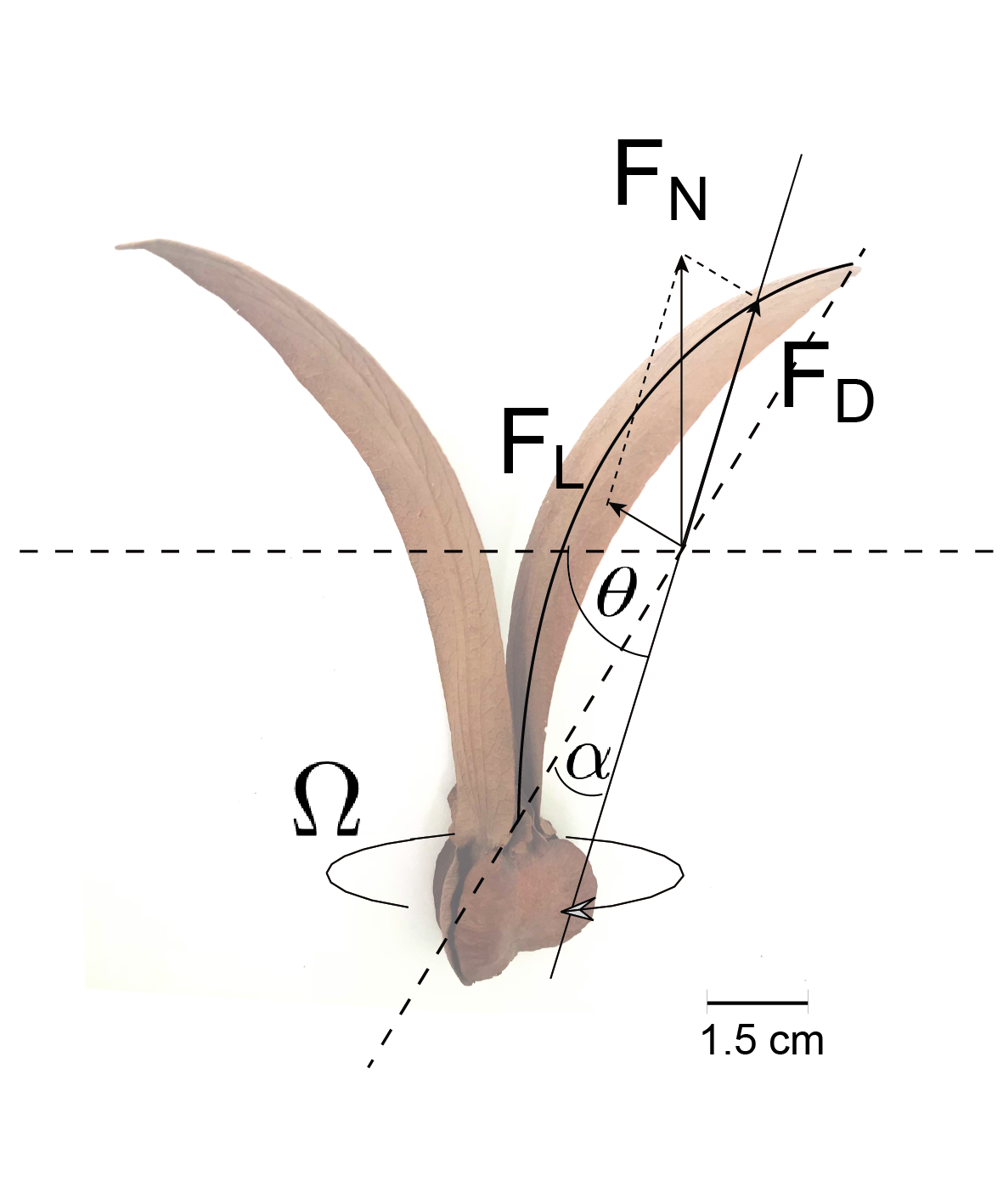}
		\caption{We parameterize the shape of the fruit and its sepals using a dried fruit and wings from the {\it Dipterocarpus alatus}. We displayed the blade element in the plane perpendicular to the wingspan. Here $\theta$ the angle of the relative wind, $\alpha$ the angle of attack, $L$ and $D$ the lift and drag components, respectively. The $F_{N}$ represents the vertical forces resulting form $F_{L}$ and $F_{D}$. The camber profile is illustrated by the curvature of the blade element. An angular
frequency is given by $\Omega = 2\pi f$, where $f$ is the rotational frequency.} \label{autora}
	\end{center}
\end{figure}
Recent investigation can be found in Refs.\cite{RAF,JRR}. As is described in Refs.\cite{RAF,JRR}, the lift force is just the component of the aerodynamic force perpendicular to the direction of the relative wind, while the drag force is the force component parallel to the wind direction. A vertical motion of the element depends on the components of the lift ($F_{L}$) and drag force ($F_{D}$) and the gravitational force ($F_{g}$) illustrated in Fig.(\ref{autora}). As suggested in Refs.\cite{RAF,JRR}, the magnitude of the lift force and the drag force on an element are proportional to the lift and drag coefficients, $C_{L}$ and $C_{D}$, respectively. We therefore assume the total force exerting on the object takes the form
\begin{eqnarray}
|{\vec F_{g}}+{\vec F}_{L}+{\vec F}_{D}| = -mg + k'_{2} v^{2},\label{force2}
\end{eqnarray}
where $v$ is the relative wind at the blade and a constant $k'_{2}$ is given by \cite{RAF,JRR}
\begin{eqnarray}
k'_{2}\equiv \frac{1}{2}\rho A(C_{L}+C_{D}),\label{kp}
\end{eqnarray}
where $\rho$ is the air density and $A$ is the area of the blade element. Specifically, the lift and drag coefficients depend on the blade element's angle of attack $\alpha$, i.e. $C_{L}(\alpha)$ and $C_{D}(\alpha)$. More specifically, along the wing segment, the lift force $F_{L}=\rho v^{2} A C_{L}(\alpha)/2$ is perpendicular to the relative wind and the drag force $F_{D}=\rho v^{2} A C_{D}(\alpha)/2$ is tangential to the direction of the wind. Regarding the classical relation, we have $C_{L}(\alpha)=2\pi \sin(\alpha)$ \cite{LDL2013}, see also Refs.\cite{RAF,JRR}, while $C_{D}(\alpha)$ can be quantified by the experiment. Using numerical values of $C_{L}+C_{D}$ given in Table \ref{table}, we can estimate the values of $C_{L}$ and $C_{D}$ separately. For instance, let us consider a sample\#1 and measure an attack angle $\alpha$. We find for this sample\#1 $\alpha\sim 20^\circ$. Using this value, we obtain $C_{L}\sim 2.15$, and then $C_{D}\sim 3.70-2.15 = 1.55$ implying that a ratio $C_{L}/C_{D}$ is about $1.4$. In the same manner, this can be implemented for other two samples. Note that our present work is focused only on the macroscopic motion of the seeds that unifies microscopic parameters, e.g. an attack angle $\alpha$. We will leave this microscopic behavior for further investigation. It is worth mentioning here that excellent detailed examinations of $C_{L}$, $C_{D}$ and $C_{L}/C_{D}$ for double-winged autorotating seeds, fruits, and other diaspores can be found in Ref.\cite{JRR}. However, the added mass of the air cannot be in principle neglected and may be added to the system for consideration. As pointed out in Refs.\cite{Norb,JRR}, this mass can be implied from the mass flow $f_m$ in unit time through the swept disk which is given by \cite{Norb,JRR} 
\begin{eqnarray}
f_{m}\equiv \rho A_{d}\Big(\frac{v+v_{f}}{2}\Big),\label{kp}
\end{eqnarray}
where $(v+v_{f})/2$ is called the through-flow velocity, $\rho$ is the air density, $A_{d}$ is the horizontally projected area swept by the wings, $v$ is the descent velocity, $v_{f}$ is the mean velocity of the air relative to the seed after passing through the area swept by the wings. In order to compare the weight of the added mass of air $(W_{m})$ with $mg$, we compute the weight via \cite{Norb,JRR}: 
\begin{eqnarray}
W_{m}=f_{m}\Big(v-v_{f}\Big)=\frac{\rho A_{d}}{2}\Big(v^{2}-v^{2}_{f}\Big).\label{kp1}
\end{eqnarray}

During the auto-rotation, the speed of the autorotating seed can be estimated using Eq.(\ref{ed1}) to obtain   $v=[1.30-1.60]\,m\,s^{-1}$ in which the terminal speed is already included as shown in Table \ref{table}. In our present work, we can grossly estimate the weight of the added mass of air by using $\rho=1 \,kg\,m^{-3}$, $A_{d}\sim 6.0\times10^{-3} \,m^{2}$, $v\sim [1.30-1.60]\,m\,s^{-1}$, and $v\gg v_{f}$
to obtain $W_{m}\sim [5.0-8.0]\times10^{-3}\,N$.
Clearly, this mass value is significantly less than the weight of the fruit $mg\sim 3\times 10^{-2}\,N$. Hence $mg/W_{m}\sim [3.7-6.0]$. From now on, it is reasonable not to consider this added mass of air. Note that numerical values of $C_{L}+C_{D}$ can be obtained by our experiments. In principle, the \lq\lq virtual mass force\rq\rq $(f_{vir})$ which is another added mass force becomes evident when we take a look at the momentum equation for the particle, see Ref.\cite{CC1998}. This force is basically proportional to the volume given by \cite{CC1998}
\begin{eqnarray}
f_{vir}\propto V_{p}\Big(\frac{Du}{Dt}-\frac{dv}{dt}\Big),\label{kp11}
\end{eqnarray}
where $V_{p}$ is the volume of the seed, and by definition
\begin{eqnarray}
\frac{D}{Dt}=\frac{\partial}{\partial t}+{\vec u}\cdot {\vec \nabla},\label{kp11}
\end{eqnarray}
and $u$ is the fluid flow velocity and $v$ is the seed velocity. However, in our case both the fluid flow velocity and the seed velocity do not change much in time and also are spatial-independent. Hence, the added-mass-force term usually proportional to a volume is approximately negligible. A straightforward application of the Newton's second law yields
\begin{eqnarray}
m\frac{d^{2}y_{2}(t)}{dt^{2}} = -mg+k'_{2}\left(\frac{dy_{2}(t)}{dt}\right)^{2},\label{str2}
\end{eqnarray}
whose general exact solutions take the form
\begin{eqnarray}
y_{2}(t) = y_{*}-\frac{m}{k'_{2}}\eln \left(\sinh \left(\frac{\sqrt{g\,k'_{2}}}{\sqrt{m}}t\right)\right),\label{ed}
\end{eqnarray}
where $y_{*}=y_{2}(t=t_{*})$. Here a coefficient $k'$ can be numerically obtained using the experimental analysis in Sec.(\ref{sec4}). Taking derivative of Eq.(\ref{ed}) with respect to time yields
\begin{eqnarray}
|v_{2}(t)| = \frac{\sqrt{g\,k'_{2}}}{\sqrt{m}}\frac{m}{k'_{2}}\coth({\cal T}),\label{ed1}
\end{eqnarray}
where we have defined ${\cal T}$ as ${\cal T}\equiv \frac{\sqrt{g\,k'_{2}}}{\sqrt{m}}t$. It is worth noting that Eq.(\ref{ed}) allows us to determine terminal velocity ($v_{\rm te}$) of the seed. The terminal velocity can be simply obtained by considering a limit where $t\rightarrow \infty$ and we find $|v_{\rm te}| = \sqrt{g\,m/k'_{2}}$ which is a constant.

\section{Experimental Setup}
\label{sec3}
\begin{figure}[H]
	\begin{center}
  		\includegraphics[width=.5\linewidth]{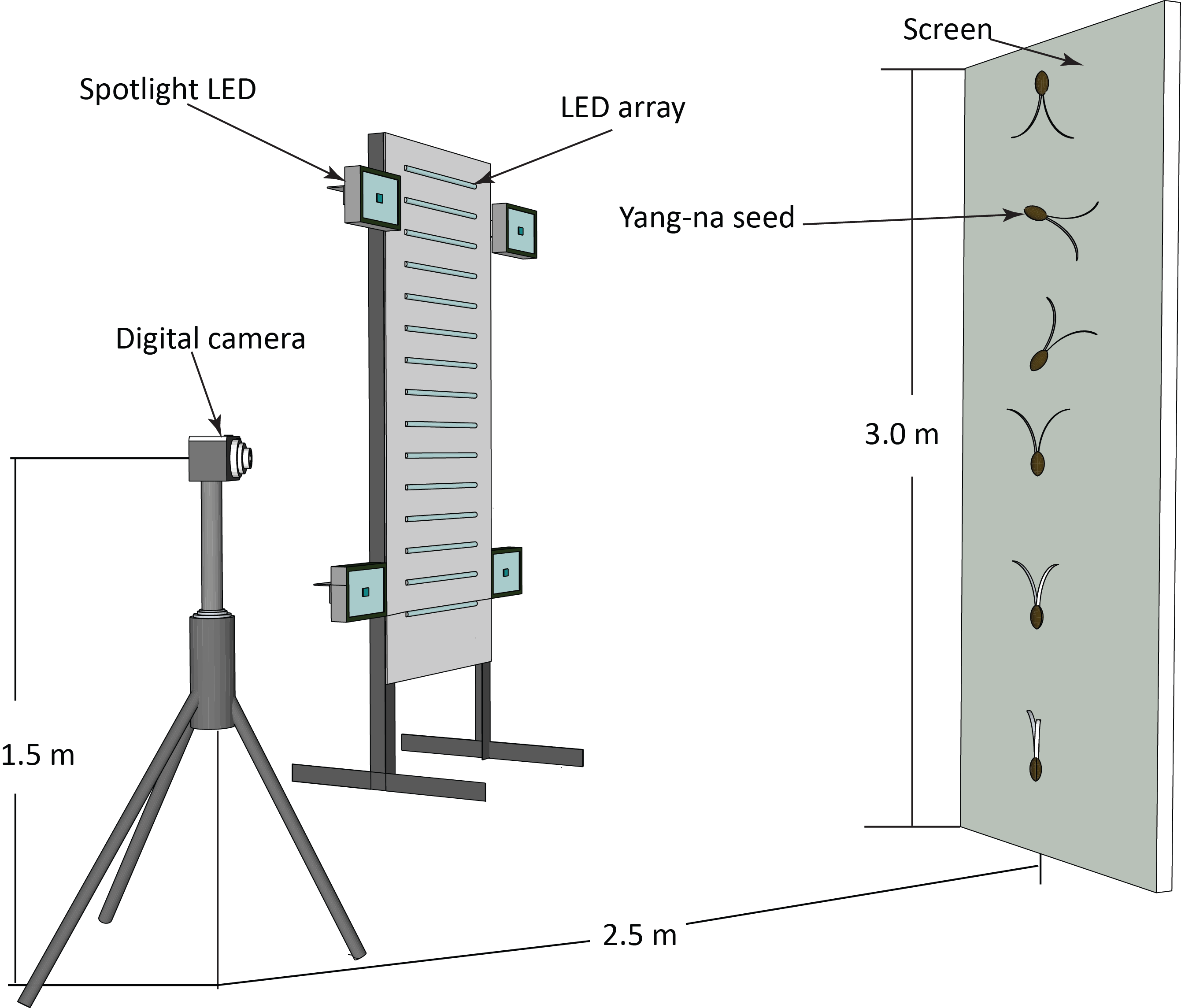}
		\caption{The schematic diagram of experimental setup. The experimental setup consists of LED spotlights, LED arrays, a digital camera, and a screen. The digital camera is used to record the falling of the Yang-na seed. The position-time data is then extracted from the recorded video using Tracker.} \label{Heff10}
	\end{center}
\end{figure}

\begin{figure}[H]
	\begin{center}
  		\includegraphics[width=.6\linewidth]{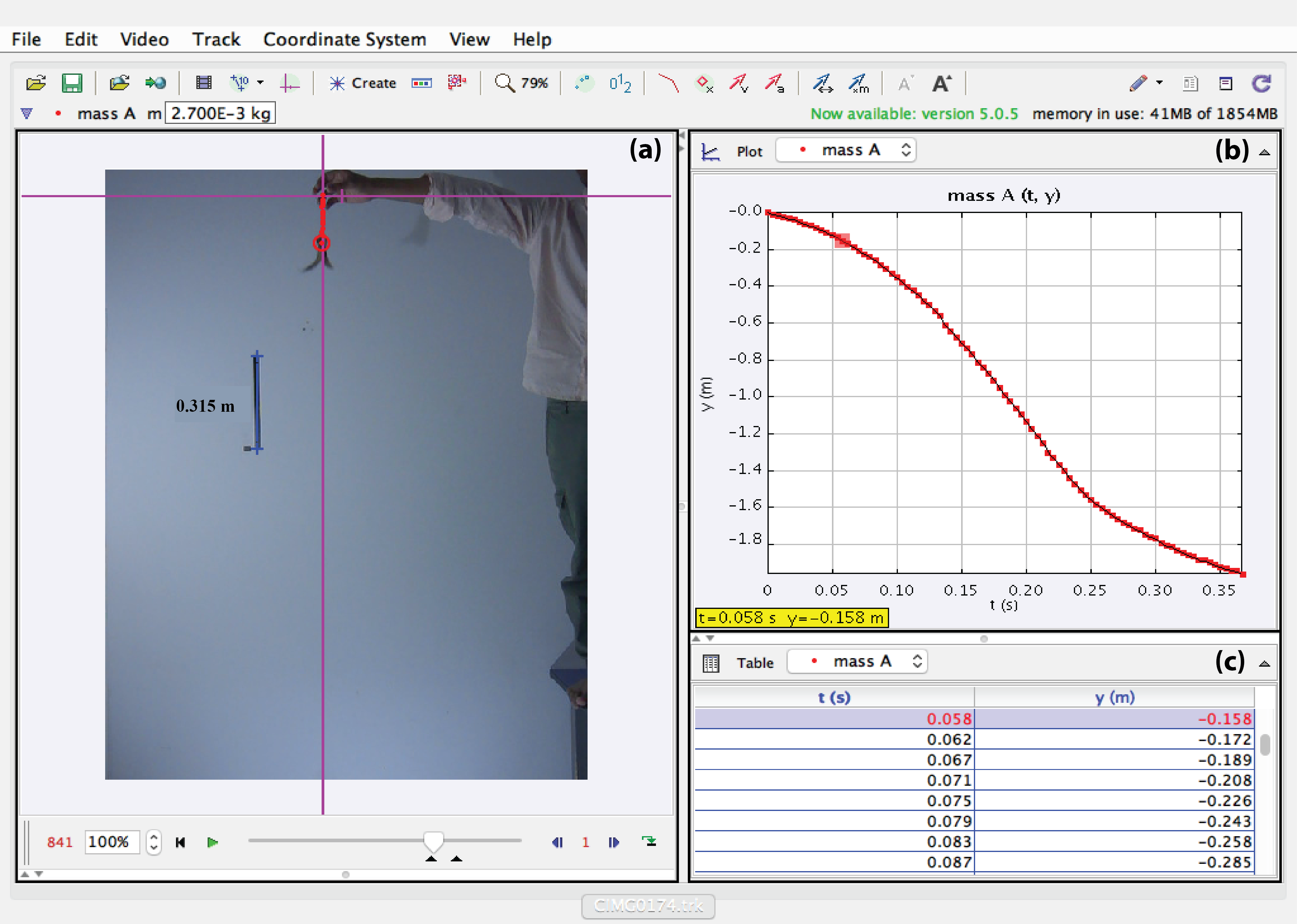}
		\caption{Tracker analysis. The analysis panel consists of 3 sections: (a) the falling of a Yang-na seed showing a cross-section reference point of origin ($x=0, y=0$) and a calibration stick of 0.315 m, (b) a plot of vertical positions of the Yang-na seed at different time points, and (c) a table of the vertical displacements at different time points.} \label{Heff14}
	\end{center}
\end{figure}

The experimental setup, as shown in Fig.(\ref{Heff10}), consists of a light panel, a high speed shutter camera: Model(EX-FH100), and a  screen. The light panel consists of 4 LED spotlights at each corner and 10 arrays of LED. The LED light source is supplied by a 12V battery to generate stationary light at 400-800 Lux. The height of the camera and distance between camera and the screen are 1.5 and 2.5 m respectively. At the beginning of the experiment, a Yang-na seed is released vertically at 3.0 m height above the floor in a close room without air flow. The digital camera is set to record in the audio-video-interleave (.avi) format at 240 frames per second (fps). The physical resolution of our video is about $9.78\times 10^{-4}$ meters per pixel. The recorded videos are calibrated within Tracker for the displacement of the falling seed. The vertical falling of three Yang-na seeds were independently tested. The vertical positions and time of each falling seed were acquired using Tracker analysis. An example of Tracker analysis is shown in the Fig.(\ref{Heff14}). The vertical positions and time of falling were further analyzed in the next section.

\section{Experimental Analysis}
\label{sec4}
Before performing any further analysis, let's summarize what we have for the background solutions and this is required to contact with the experiment. We use the experimental data to determine all unknown parameters that we demonstrate below. We conclude the predicted solutions as follows:
\begin{equation}
    y(t) = \begin{cases}
        y_{1}(t_{1})=y(t)|_{A\rightarrow B}=y_{*,1}-\frac{m}{k'_{1}} \eln \left(\cosh \left(\frac{\sqrt{g\, k'_{1}}}{\sqrt{m}}t_{1}\right)\right) & \text{for } t_{0}\leq t_{1} \leq t_{*} , \\
       y_{2}(t_{2})=y(t)|_{B\rightarrow C}= y_{*,2}-\frac{m}{k'_{2}}\eln \left(\sinh \left(\frac{\sqrt{g\,k'_{2}}}{\sqrt{m}}t_{2}\right)\right) & \text{for } t_{*} \leq t_{2} \leq t\,,
        \end{cases}
        \label{sum}
  \end{equation}
where $t_{*}$ can be fixed from our experimental results. It is worth noting that the displacement is dependent on not only mass but also the geometry of the seed. We explicitly observed in Eq.(7) that the displacement is dependent on the parameter $k_{2}'$. This parameter unifies a macroscopic properties implying that the displacement is also dependent on the geometry of the seed. More concretely, values of such a parameter are proportional to the area of the blade element $(A)$, and the blade element’s angle of attack $(\alpha)$. In the same manner, this is true for $k_{1}'$. Interestingly, we found that our predicted solutions are in agreement with the experiment. To be more precise, the whole trajectory of the falling seed of Yang-na can be completely parametrized by our solutions given in Eq.(\ref{sum}). 
\begin{table*}[ht!]
 \begin{center}
  \begin{tabular}{ccccccc}
  \hline
Sample & Mass & Wing length & Wing area & $(k'_{1},k'_{2})$ & $C_{L}+C_{D}$ & Terminal velocity (best-fitting value) \\
 & $g$ & $\cdot10^{-1}\,m$ & $\cdot10^{-3}\, m^{-2}$ & $\cdot 10^{-3}$ &  & $m\cdot s^{-1}$ \\
  \hline\hline
  1 & 2.70 & 0.95 & $6.24$ & $(0.93,14.14)$ & $3.70$ & 1.37\\
  \hline
  2 & 2.85 & 1.04 & $6.79$ & $(1.36,11.80)$ & $2.84$ & 1.54 \\
  \hline
 3 & 3.20 & 1.00 & $6.35$ & $(1.49,13.00)$ & $3.34$ & 1.55\\
  \hline
  \end{tabular}
  \caption{Parameters of the Yang-na seeds are used in the present analysis. Note that numerical values of $(k'_{1},k'_{2})$ in the last column are obtained from best fitting Eq.(\ref{sum}) with the experimental data. Note that wing areas of the {\it Dipterocarpus alatus} seed were calculated by summing the product of length and width of their two wings \cite{JR}. In order to obtain numerical values of $C_{L}+C_{D}$, we have used air density of approximately $1.225\, kg/m^{3}$ according to ISA (International Standard Atmosphere).}
  \label{table}
 \end{center}
 \end{table*}
The greater the drag force is, the faster Yang-na seed will reach its terminal velocity. Regarding the term of lift force $k'_{2}v^2$, the three Yang-na seeds display approximately similar values of $k'_{2}$. The seeds take about 0.5-0.7 second to achieve its terminal velocity in the auto-rotation flight of the second interval.
\begin{figure}[H]
	\begin{center}
        \includegraphics[width=1.0\linewidth]{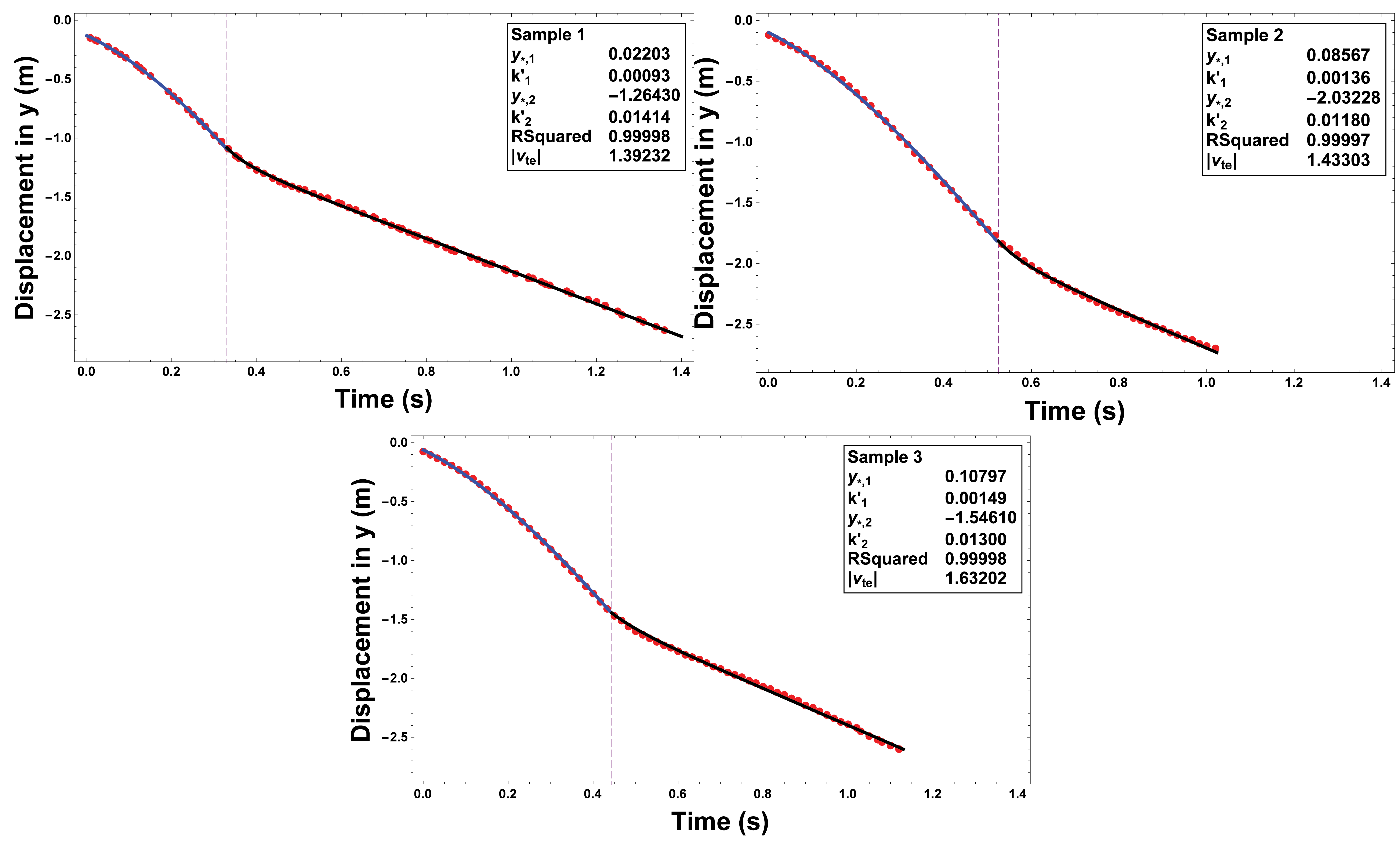}
		\caption{The plot of experimental data (red dot) fitted with predicted vertical displacement (line) of the three Yang-na seed Samples. Vertical dash line represent a transition of seed from a falling$\&$flipping interval to auto-rotation one corresponding to $t_{*}$. Remarkably, the analysis of the displacement was considered after the initial point. Therefore the plot does not start at $(0,0)$.} \label{Heff200}
	\end{center}
\end{figure}

Moreover, we observe that $k'$ pretty much depends on $A(C_{L}+C_{D})$. From Table (\ref{table}), we find that the values of $k'$ are not significantly distinct among all three seeds.

\section{Concluding remarks}

We performed analytical and experimental investigation of the vertical motion of {\it Dipterocarpus alatus} seed, locally so-called Yang-na. In this work, we assumed the drag forces exerting on the Yang-na seeds depend on the velocity. We figured out the equation of motion and analytically solved to obtain the exact solutions. We determined the terminal velocity to obtain $|v_{\rm te}| = \sqrt{g\,m/k'_{2}}$. Interestingly, we found that our predicted solutions are in agreement with the experiment. As a result, the whole vertical trajectory of the falling seed of Yang-na can be completely parametrized by our predicted solutions. The perfect agreement between our predicted solutions and experimental data has been illuminatingly displayed in Fig.(\ref{Heff200}).

We observed that the lift force generated in the wing-seed system is proportional to the velocity squared. As a result, our framework constitutes the complete template for vertical motions of any seed with an arbitrary number of wings which can be effectively described by our predicted solutions given in Eq.(\ref{sum}). To be more precise, the vertical behavior of such wing seeds could be completely satisfied by our exact solutions. A straightforward verification can be directly done by following our scenario. In addition, regarding the Yang-na motion, one can examine rotational motion by following the framework done by Ref.\cite{JR} for different types of wing seeds to examine the angular speed. We strongly believe that our results can basically be generalized to vertical flight model of all wing seeds with external wind. However, this requires further experimental validation. In addition, the present work initiates the extension to broad circumstances, e.g. physics education. Noticeably, our model completely facilitates a high predictable nature of a Yang-na's vertical motion and can be translated to the rehabilitation and sustainable management of the Yang-na population to ameliorate the forest decline.

Our highlight is threefold: (1) The studies related to the vertical motion of flying seed have been investigated so
far. However our work is the first achievement which completely explain entire flight sequence. In addition, we demonstrate that theory and experiment can simply be reconciled. We employ
mathematical model based on a simple Newton’s law to unravel the nature of { \it Dipterocarpus alatus}
seed. We mathematically and physically proves that the seed of { \it Dipterocarpus alatus} travels vertically
straight downward without deflection. Having assumed the center of gravity always situated along the
vertical axis, we derive equations of motion to obtain the exact solution which perfectly agree with the
experimental observation. This could be an explanation supporting the poor dispersion nature of
{ \it Dipterocarpus alatus } seed. The terminal velocity can be theoretically predicted. Moreover, our mathematical model can be generalized to describe any type
of wing seeds. (2) We believe that our achievement will be valuable to the large community of
STEM/STEAM education to promote an understanding in the topic integrating mathematics, physics,
biology, art and technology. Our framework constitutes learning model to improve the ability of creative thinking, analytical thinking and problem solving skills on the concept of forces and motion
applicable from high school to college levels and (3) Our model completely facilitates a high
predictable nature of a { \it Dipterocarpus alatus } seed’s vertical motion and can be translated to the
rehabilitation and sustainable management of the seed population to ameliorate the forest decline.

\section*{Acknowledgments}
This work is financially supported by the National Research Council of Thailand (NRCT) with grant number 61119. We thank Ms.Kanitha Kongoop for collaborating at an early stage of this work. This research was partially supported by the New Strategic Research (P2P) project with grant number CGS-P2P-2562-040, Walailak University, Thailand.


\section*{References}
\begin{thebibliography}{99} 

\bibitem{VL} 
 V~Ly, K~Nanthavong, R~Pooma, M~Barstow, H~T~Luu, E~Khou, and M~F~Newman {\it IUCN} 2017 e. T33007A2829912 (2017).

\bibitem{TS} 
T~Smitinand, T~Santisuk, and C~Phengklai {\it Thai For. Bull.} 0 133 (1970).

\bibitem{KT}
K~Tennakone {\it J. Natl. Sci. Found. Sri Lanka} 45 201 (2017).

\bibitem{IL} 
I~Lee and H~Choi {\it Phys. Rev. Fluids} 2 090511 (2017).

\bibitem{AC} 
A~V~C~Camposano, N~C~Virtudes, R~E~S~Otadoy, and R~Violanda {\it IOP Conf. Ser. Mater. Sci. Eng.} 79 1 (2015).

\bibitem{SS} 
 S Sasaki {\it Proc. Japan Acad. Ser. B Phys. Biol. Sci.} 84 31 (2008).

\bibitem{JR} 
J~R~Smith, R~Bagchi, J~Ellens, C~J~Kettle, D~F~R~P~Burslem, C~R~Maycock, E~Khoo, and J ~Ghazoul {\it Ecol. Evol. } 5 1794 (2015).

\bibitem{SA} 
S~Appanah and J~M~Turnbull {\it A review of dipterocarps: taxonomy, ecology and silviculture} (1998).

\bibitem{CL} 
C~Ladera and P~Pineda {\it Latin-American J. Phys. Educ. } 3 557 (2009).

\bibitem{DB} 
D~Brown, Tracker. Video Analysis and Modeling Tool, https://physlets.org/tracker/ .

\bibitem{TE}
T~Eadkhong, R~Rajsadorn, P~Jannual, and S~Danworaphong {\it Eur. J. Phys.} 33 615 (2012).

\bibitem{LKW}
L~K~Wee, C~Chew, G~H~Goh, S~Tan, and T~L~Lee {\it Phys. Educ.} 47 448 (2012).

\bibitem{JK}
J~Kinchin {\it Phys. Educ.} 51 053003 (2016).

\bibitem{DUS}
D~U~Suwarno {\it J. Sci. Sci. Educ.} 1 75 (2017).

\bibitem{JDbook} 
J.~D.~Anderson, {\it Fundamentals of aerodynamics}, Fifth Edition, McGraw-Hill series in aeronautical and aerospace engineering, page 296-297

\bibitem{KV} 
K~Varshney, S~Chang, and Z~J~Wang {\it Nonlinearity } 25 C1 (2012).

\bibitem{Norb} 
R~A~Norberg {\it Biol. Rev. Camb. Philos. Soc. } 48 561 (1973).


\bibitem{Aki} 
A~Azuma and K~Yasuda {\it J. Theor. Biol. } 138 23 (1989).

\bibitem{Lee} 
S~J~Lee, E~J~Lee, and M~H~Sohn {\it Exp. Fluids} 55 1718 (2014).

\bibitem{DL2009} 
D.~Lentink, W.~B.~Dickson, J.~L.~van Leeuwen, M.~H.~Dickinson, {\it Leading-Edge Vortices Elevate Lift of Autorotating Plant Seeds}, Science 324 (2009) 5933, pp.1438-40   

\bibitem{JS2019} 
J.~S.~Carlton FREng, {\it Cavitation}, in Marine Propellers and Propulsion (Fourth Edition), 2019

\bibitem{JJ2019}
J.~Jeong and F.~Hussain, {\it On the identification of a vortex}, J. Fluid Mech. 285 (1995) 69

\bibitem{TM1979}
T.~Maxworthy, {\it Experiments on the Weis-Fogh mechanism of lift generation by insects in hovering flight}. Part 1. Dynamics of the “fling”, Journal of Fluid Mechanics, 93 (1979) 01, 47

\bibitem{RAF} 
R.~A.~Fauli, J.~Rabault, and A.~Carlson {\it Phys. Rev. E} 100 013108 (2019).

\bibitem{JRR}
J.~Rabault, R.~A.~Fauli, and A.~Carlson {\it Phys. Rev. Lett.} 122 024501 (2019).

\bibitem{LDL2013}
L.~D.~Landau and E.~M.~Lifshitz, {\it Fluid mechanics} (Elsevier, Amsterdam, 2013), Vol.6.

\bibitem{CC1998}
C.~Crowe, T.~Clayton, M.~Sommerfeld, Y.~Tsuji (1998). {\it Multiphase flows with droplets and particles}, CRC Press. p.81. ISBN 978-0-8493-9469-0

\end{thebibliography}
\end{document}